# Modelization of flow electrification in a polymer melt.

F Flores[1], D Graebling[1], A Allal[1] and C Guerret-Piécourt[2]
[1]Laboratoire de Physico-Chimie des Polymères, UMR CNRS 5067, CURS, Avenue de l'université, BP 1155, 64013 Pau cedex France
[2]Laboratoire de Tribologie et Dynamique des Systèmes, UMR CNRS 5513, Ecole Centrale de Lyon, 36 avenue Guy de Collongue, 69134 Ecully cedex France

E-mail : christelle.guerret@ec-lyon.fr

**Abstract**:
Flow electrification of polymer melts is an important side effect of polymer processing. The studies dealing with this phenomenon are seldom and most of the scientific work has been focused on flow electrification of aqueous and insulating Newtonian liquids. From that prior art it is well established that the flow electrification in Newtonian liquids is a consequence of the formation of an ionic double layer. Convection of this layer induces the electrification of the liquid at the outlet of the pipe. In those models, the key parameters governing the flow electrification are thus the intrinsic electrical properties of the polymer and the flow characteristics. In this work, we reconsider the assumptions made previously and we propose a new approach to modelise the flow electrification in the particular case of non-Newtonian polymer materials in laminar flow conditions. We establish that, a key parameter for the electrification quantification in the polymer melt is the shape of the velocity profile. Additionally, in some cases, we show that a slip velocity at the polymer/die wall interface must be considered to describe accurately the electrification. As a consequence, we deduce that the slip velocity at the interface can be calculated by measuring the electrification: this work gives an alternative manner to measure the slip velocity during polymer flow.

**1. Introduction**

Flow electrification has been extensively studied since the second World War due to its economical and theoretical interest [1-3]. On one hand, many issues are related to flow electrification like fire or explosions particularly in the petroleum industry. On the other hand, flow electrification can also generate trapped space charges in an extruded material. Fortunately, many practical solutions have been found in each case to improve safety but the electrification mechanism is still partially unknown.
In many papers, scientists have developed electrification theories for liquids [1, 4-7]. These theories explain the electrification phenomenon through the double layer theory. The electrical double layer is due to ions currents (diffusion, convection and migration) generated by electrochemical processes at the wall [1, 8]. In these models, two main components are used for the calculation of the generated electric charge: the ionic charge density and the fluid velocity profiles in the pipe. For standard liquids, like aqueous solutions or oils, the authors used Newtonian velocity profiles in laminar conditions.
Contrary to the numerous investigations on the flow electrification of Newtonian liquids, to our knowledge only one study has considered the particular case of melt polymers in laminar conditions [5, 6]. In this paper, Taylor explained the electrokinetic charging of polymers by the classical theory of the double layer formed at the polymer/die interface during the capillary extrusion. However polymers are non Newtonian materials and Taylor then used a specific fluid velocity profile to describe this non Newtonian behavior [5]. This specific velocity profile resulted from the Ostwald power law that links the wall shear stress and the shear rate applied to the flowing polymer. Starting from Taylor's investigations, the basic concept of the new approach exposed in this work is to utilize a non Newtonian velocity profile appropriate to the polymer behavior. Indeed Taylor's velocity profile does not correctly describe the polymer behavior in the whole range of the shear rates characteristics of extrusion. The innovation of the present work is then to employ a velocity profile more adapted for polymers than the Ostwald ones. We show that this careful choice of profile has a strong influence on the generated tribocharging amount. Perez-Gonzalez et al. have also proposed one easiest model but they didn't give any physical explanations of the phenomenon [9].

Starting from the description of the double layer building, we first show that the usual approximations used in the resolution of the charge transport equations are not valid in the polymer case due to their intrinsic electrical properties. Concerning the second component of the electric charge evaluation, non Newtonian velocity profiles are then considered and their influence on the electrification is discussed. The experimental measurements of charges are compared with the theoretical predictions in order to validate the model and to put in evidence some discrepancies. We demonstrate that these discrepancies can be solved by introducing a wall slip velocity in the velocity profile. Furthermore, one can note that the knowledge of polymer slip behavior is very important for the understanding of polymer flow properties [10].

## 2. Experimental details

### *2.1. Experimental set-up*
The experiments on flow electrification of molten polymers were performed using a capillary rheometer coupled with a fieldmeter located under a Faraday collector. This experimental set-up is close to the ones used by Taylor and it has been fully described in previous works [5, 10, 11]. Opposite to other set-ups it allows to measure accurately small amounts of electrical charges: the lowest level of measured electric charges is some tens of picocoulomb, with a resolution of 1 pC [12]. Polymers are extruded with a Rheoflixer capillary rheometer at imposed flow rate. The molten polymer flows through tungstene carbide capillaries with different length over diameter ratios (L/D: 30/1 and 40/0.5). Experiments were performed at two temperatures 80°C and 60°C.

### *2.2. Materials*
Two commercial polydimethylsiloxanes (PDMS A and B) were studied in this work. As shown later the electrification calculation requires the precise knowledge of the viscosity behavior of the studied polymers. So, the flow curves (viscosity vs shear rate curves) have been obtained thanks to a rheometer at imposed strain (Rheometric Scientific RDA II). Polymers have a non Newtonian behavior because of their structure of long macromolecular chains are entangled or not. For many polymers these entanglements lead to a shear-thinning behavior at high shear rates. Figure 1 presents the flow curve of the PDMS B at T = 60°C.

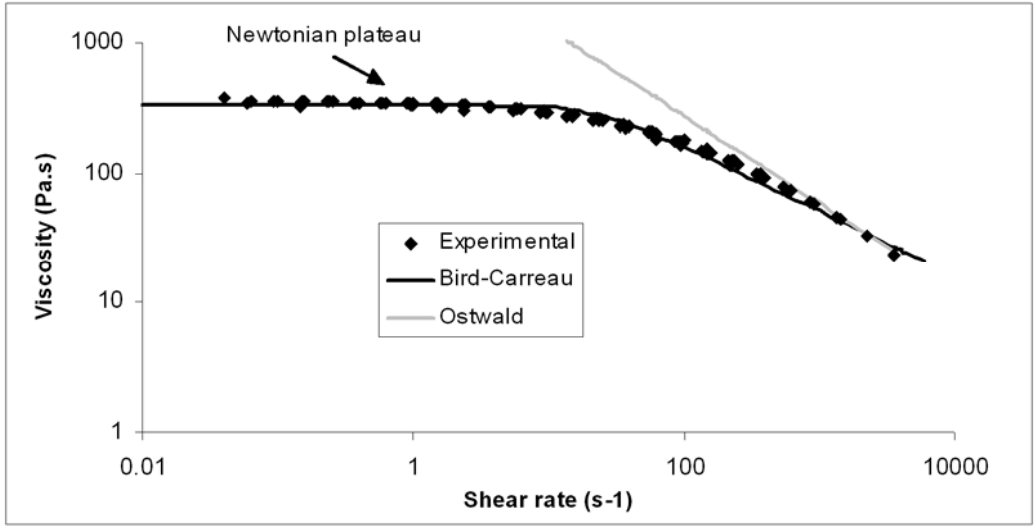

**Figure 1.** Viscosity versus shear rate pour PDMS B at T = 60°C

In figure 1, one can observe two parts on the experimental curve. The first one is the Newtonian plateau where the viscosity is independents of the shear rate (below 10 s$^{-1}$). In the second part, above 10 s$^{-1}$, the viscosity decreases with the shear rate. Many laws have been established to modelise this shear-thinning and we chose two of them. The first one is the classical Ostwald law used by Taylor [5]:

$$\eta = \frac{\sigma_s}{\dot{\gamma}} = K\dot{\gamma}^{n-1} \quad (1)$$

where $\sigma_s$ is the shear stress and $\dot{\gamma}$ is the shear rate. K, the consistence and n, the pseudo plastic index are constant (K and n are obtained by fitting on the experimental data). It can be shown in figure 1 that the Ostwald law gives good agreements with the experimental results at high shear rates (above 100 s$^{-1}$) but for low shear rates significant differences can be observed. On the opposite, other laws like the well-known Bird-Carreau law gives best fit in all Newtonian shear rate region [13]:

$$\eta = \eta_\infty + (\eta_0 - \eta_\infty)\left[1 + (\lambda\dot{\gamma})^2\right]^{\frac{n-1}{2}} \quad (2)$$

where η is the viscosity, $\eta_0$ is the viscosity on the Newtonian plateau, $\eta_\infty$ is the viscosity at infinite shear rates ($\eta_\infty = 0$ in the case of polymer melts), λ is the relaxation time, $\dot{\gamma}$ is the shear rate and n a pseudo plastic index. In figure 1, one can observe the good correlation between the experimental values and the Bird-Carreau law for low and high shear rates in the case of PDMS B at T = 60°C.

Note that the Bird-Carreau law (equation 2) is well adapted for several polymers. But its parameters must be modified for each case and they mainly depend on the polymer features and on the temperature. The non Newtonian behaviors are quite similar for PDMS A and B used either at T = 60°C or at T = 80°C. The thermodependence of the viscosity of polymers can be describe by the Arrhenius' law or Andrade-Eyring's law:

$$\eta(T) = \eta_{ref} \exp\left(\frac{\Delta E}{R}\left(\frac{1}{T} - \frac{1}{T_{ref}}\right)\right) \quad (3)$$

where $\eta_{ref}$ is the viscosity of the polymer at the reference temperature, ΔE is the activation energy and R the ideal gas constant. In the case of polydimethylsiloxane. ΔE is equal to 10,8 kJ/mol.
Fit parameters for PDMS A and B at both temperatures are reported in Table 1.

**Table 1.** Bird-Carreau coefficients with $\eta_\infty=0$

| Polymer | Temperature (°C) | $\eta_0$ (Pa.s) | λ (s) | n |
|---|---|---|---|---|
| A | 60 | 160 | 0.02 | 0.6 |
|   | 80 | 129 | 0.01 | 0.6 |
| B | 60 | 330 | 0.04 | 0.5 |
|   | 80 | 217 | 0.04 | 0.5 |

From table 1, one can see that the value of the Newtonian plateau change with the temperature and the polymer type. Polymer B has a viscosity higher than polymer A for the two temperatures (60°C and 80°C). Another important feature is the decrease of the viscosity with an increase of the temperature for each polymer A and B.

The used industrial polymers contain some additive impurities (anti oxidative, processing aid, etc...) or polymerization residues. In all the cases, some anions and cations exist in the liquid polymers and their conductivities can be measured. The electrical conductivities for the polymers studied in this work have been determined thanks to a LCM-8716 conductimeter from Alff engineering. The protocol used for the conductivity measurement is the following: (i) because polymers are highly viscous materials, one night is necessary to obtain a good filling of the test cell before each measurement; (ii) the polymer melt in the test cell is then excited with low amplitude (30 V) low frequency (0.5 Hz) alternate square wave voltage. (iii) Thanks to current measurements, the capacitance and the conductance can be determined and values of the relative permittivity and volume conductivity are given by the apparatus. Results are reported in table 2.

**Table 2.** Electrical conductivity and permittivity of PDMS

| Polymer | Temperature (°C) | Relative permittivity | Conductivity (S/m) |
|---------|------------------|----------------------|--------------------|
| A       | 60               | 2.64                 | $1.17 \cdot 10^{-10}$ |
|         | 80               | 2.62                 | $5.5 \cdot 10^{-9}$  |
| B       | 60               | 2.62                 | $2.35 \cdot 10^{-12}$ |
|         | 80               | 2.54                 | $9.75 \cdot 10^{-12}$ |

For each polymer, one can remark an increase of the conductivity with the increasing temperature. Note also the decrease of the conductivity from polymer A to polymer B. This difference of conductivity may be due to different ionic concentrations in these polymers or to the increase of the polymer viscosity from PDMS A to PDMS B.

## 3. Model

One first attempt to explain flow electrification phenomenon was made by Gavis and Koszman [2]. A streaming current is always associated with the flow of liquids through capillaries. This tribocharging is caused by the liquid free electric charges, i.e. ions, due to additives and impurities existing in the liquid or in the polymer melt [1, 2, 14]. Electrochemical processes near the capillary wall induce the development of an electrical double layer. In the Stern model, the double layer consists in two regions, the compact layer and the diffuse layer as described in figure 2.

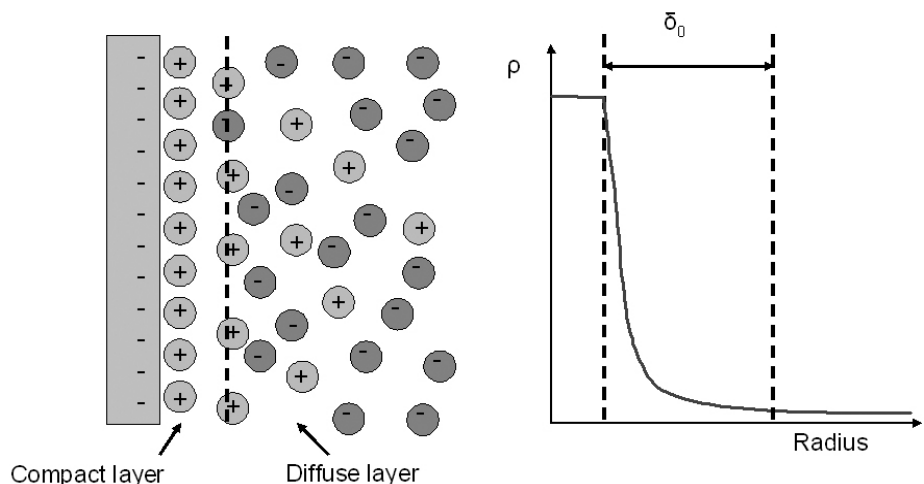

**Figure 2.** Schematic diagram of an electrical double layer and simplified representation of charge distribution with ρ the electrical charge density and $\delta_0$ the Debye length.

The compact layer is about one or two ionic diameters thick close to the capillary wall but the diffuse layer is not confined to the polymer/die interface and its thickness is given by the Debye length $\delta_0$ expressed as follow [1]:

$$\delta_0 = \sqrt{\frac{\varepsilon D}{\sigma}}, \tag{4}$$

where $\varepsilon$ and $\sigma$ are respectively the electrical permittivity and conductivity and D the ionic diffusion coefficient. This diffusion coefficient in polymers can be estimated from the mobility b [14, 15]:

$$D = \frac{bRT}{zF}, \tag{5}$$

where R is the gas constant, T is the temperature, z is the ion charge and F is the Faraday constant. The mobility is given by:

$$b = \frac{ze}{6\pi\eta r_{ionic}}, \tag{6}$$

where e is the charge of the electron, $\eta$ is the viscosity of polymer liquid and $r_{ionic}$ is the ionic radius (about $1.10^{-10}$m). Harvey et al. used these relations (5) and (6) in the case of ionic diffusion in insulating oils but Brochart-Wyart et al. showed that, in the case of ultrafine particles (in comparison with polymer chain size), such relations can be also used for ionic diffusion in polymer materials [14, 15]. Note that in the particular case of polymer, the viscosity will change with the shear rate and diffusion coefficient must be re-calculated for each shear rate considered in this work.

Under forced convection, the diffuse layer is exposed to the fluid flow and a current is generated. To determine the charge density generated at the outlet of the die, it is necessary to determine the charge profile due to the double layer formation inside the die and the velocity profile linked to the fluid flow. When these two profiles are known, the total charge convected by the flow is given by,

$$Q = \frac{2}{R_d^2 V_{mean}} \int_0^{R_d} \rho(r) \times V(r) r\, dr \tag{7}$$

where $R_d$ is the capillary radius, $V_{mean}$ is the mean velocity, $\rho(r)$ is the charge density at radius r and $V(r)$ is the shape velocity in the capillary.

*3.1. Build-up of the differential equations governing the charge distribution*

In this study, we use the formalism initially established by Touchard [16]. It is important to recall the whole calculations because the approximations generally used in the resolution of the charge transport equations are not valid in the polymer case as shown in the next paragraph. In liquids, the transport of charges occurs by three mechanisms : diffusion (8), migration (9) and convection (10) described as follow [1, 2]:

$$\vec{J}_{Di} = -D_i \overrightarrow{grad}(n_i), \tag{8}$$

$$\vec{J}_{Mi} = -\frac{n_i e z_i D_i}{kT} \overrightarrow{grad}\,\Psi, \tag{9}$$

and

$$\vec{J}_{Ci} = n_i \vec{V} \tag{10}$$

where, $D_i$, $n_i$ and $z_i$ are respectively the diffusion coefficient, the number and the charge of an ion i, k is the Boltzmann constant and $\vec{V}$ the velocity.

The relation between the charge density $\rho$ and the electrical potential $\Psi$ is obtained thanks to Poisson's equation:

$$\Delta\Psi = -\frac{\rho}{\varepsilon} \tag{11}$$

Total current for the ion i is given by:

$$\vec{J}_i = -D_i \overrightarrow{grad}(n_i) - n_i \frac{e z_i D_i}{kT} \overrightarrow{grad}\,\Psi + n_i \vec{V} \tag{12}$$

With the simplifying assumptions that the diffusion coefficient of all anions and cations are equal, $D_i = D$, and that the charges of cations and anions are the same $z_i = z$ for cations and $z_i = -z$ for anions, one easily obtains the current densities for the cations P (13) and the anions N (14):

$$\vec{i}_P = -ezD \, \vec{grad} \, n_P - \frac{e^2 z^2 D n_P}{kT} \vec{grad} \, \Psi + ez n_P \vec{V} \tag{13}$$

and

$$\vec{i}_N = ezD \, \vec{grad} \, n_N - \frac{e^2 z^2 D n_N}{kT} \vec{grad} \, \Psi - ez n_N \vec{V} \tag{14}$$

where $n_P$ is the number of cations, $n_N$ is the number of anions. In this study, the ions involved in the double layer are still unknown so the particular case of symmetric z:z "electrolyte" case will be considered. The diffusion of small particles that are supposed to be the more mobile in polymers can be evaluated from equation (5). As we assume that anionic and cationic charges and radii are the same, then the diffusion coefficient of both anions and cations are also supposed to be equal. Note that these simplifying assumptions on the equality of the diffusion coefficients and ionic charges are usually performed in other works and lead to good correlations with experimental results [1, 7].

Then, the total current density and the conjugated total current density may be expressed as follow,

$$\vec{i} = ezD \left( \vec{grad}(n_N) - \vec{grad}(n_P) \right) - \frac{e^2 z^2 D}{kT} (n_P + n_N) \vec{grad} \, \Psi + ez(n_P - n_N) \vec{V} \tag{15}$$

and

$$\vec{i}^* = -ezD \left( \vec{grad}(n_N) + \vec{grad}(n_P) \right) - \frac{e^2 z^2 D}{kT} (n_P - n_N) \vec{grad} \, \Psi + ez(n_P + n_N) \vec{V} \tag{16}$$

Considering the expressions of the space charge density ρ, and the bulk conductivity σ as defined by Touchard:

$$\rho = ez(n_P - n_N) \tag{17}$$

and

$$\sigma = \frac{e^2 z^2}{kT} D(n_P + n_N) \tag{18}$$

(17) and (18) become:

$$\vec{i} = -D \, \vec{grad} \, \rho - \sigma \, \vec{grad} \, \Psi + \rho \vec{V} \tag{19}$$

and

$$\vec{i}^* = -\frac{kT}{ez} \vec{grad} \, \sigma - \frac{ezD}{kT} \rho \, \vec{grad} \, \Psi + \frac{kT}{ezD} \sigma \vec{V} \tag{20}$$

One important parameter governing the flow electrification phenomenon is the development time of the electrical double layer [17]. Indeed, this time is controlled by the wall physico-chemical reaction time, by the charge relaxation time, and by the time spent in the die by the melt polymer. If this last time is longer than the two others, the equilibrium conditions are reached, and the charge profile reaches its steady value. There is no net transfer of charge once the double layer has formed. In that case one considers that the double layer is fully developed. The time necessary to obtain this fully developed double layer depends on ionic diffusion, conductivity and permittivity and is discussed later (equation (36)). One can then write $\vec{i} = \vec{i}^* = \vec{0}$ and $\vec{V} = \vec{0}$. Equations (19) and (20) become:

$$D \, \vec{grad} \, \rho + \sigma \, \vec{grad} \, \Psi = \vec{0} \tag{21}$$

and

$$-\frac{kT}{ez} \vec{grad} \, \sigma - \frac{ezD}{kT} \rho \, \vec{grad} \, \Psi = \vec{0} \tag{22}$$

In cylindrical coordinates, for cylindrical pipes or capillaries, equations (11), (21) and (22) become:

$$\frac{d^2 \Psi}{dr^2} + \frac{1}{r} \frac{d\Psi}{dr} = -\frac{\rho}{\varepsilon} \tag{23}$$

$$D\frac{d\rho}{dr} + \sigma\frac{d\Psi}{dr} = 0 \tag{24}$$

$$-\frac{kT}{ez}\frac{d\sigma}{dr} - \frac{ezD}{kT}\rho\frac{d\Psi}{dr} = 0 \tag{25}$$

When $\Psi$ is eliminated between equations (24) and (25) an equation for $\sigma$ is obtained by assuming that the temperature within the die is constant:

$$\sigma^2 = \sigma_0^2 + \rho^2\frac{D^2 e^2 z^2}{k^2 T^2} \tag{26}$$

with the conditions concerning the electrical conductivity $\sigma=\sigma_0$ and the space charge density $\rho=0$ at radius r=0. If equation (26) is reported in equation (24), one then obtains,

$$D\frac{d\rho}{dr} + \sqrt{\sigma_0^2 + \frac{D^2 e^2 z^2}{k^2 T^2}\rho^2}\frac{d\Psi}{dr} = 0 \tag{27}$$

*3.2. Resolution of the equation in the case of polymers*

To solve these equations, two extreme cases are generally considered depending on the conductivity of the liquid and on the extent of the double layer. Generally, in the case of liquids such as hydrocarbons, authors make the assumption that the term including the space charge density $\frac{D^2 e^2 z^2}{k^2 T^2}\rho^2$ is weak compared to the conductivity at the die center $\sigma_0$ [17]. This hypothesis is generally made because it induces for hydrocarbons a low error (few percents) and it allows finding an analytical solution.

In that case, the simplifying assumption $\sigma_0^2 \gg \frac{D^2 e^2 z^2}{k^2 T^2}\rho^2$ is made, and equation (27) becomes [1, 17, 18],

$$D\frac{d\rho}{dr} + \sigma_0\frac{d\Psi}{dr} = 0 \tag{28}$$

When $\Psi$ is eliminated between equations (28) and (23), one obtains an equation for $\rho$ as a function of the position only

$$\frac{d^2\rho}{dr^2} + \frac{1}{r}\frac{d\rho}{dr} - \frac{\sigma_0}{D\varepsilon}\rho = 0 \tag{29}$$

Then, as demonstrated by Touchard, the space charge density may be expressed as follow [1]:

$$\rho(r) = \rho_w\frac{I_0\left(\frac{\alpha r}{\delta_0}\right)}{I_0\left(\frac{\alpha R_d}{\delta_0}\right)} \tag{30}$$

where $\rho_w$ is the wall space charge density, $R_d$ is the capillary radius, $\alpha$ a coefficient close to one which takes into account the ionic diffusion coefficient and $I_0$ is the modified Bessel function of zero order. The wall space charge density can not be determined by direct measurement. But it could be evaluated from the electrification measurements [17].

On the contrary, if no simplifying assumption is made, $\Psi$ is eliminated between equations (27) and (23), one then obtains:

$$\frac{d^2\rho}{dr^2} - \frac{B}{\sigma_0^2 + B\rho^2}\rho\left(\frac{d\rho}{dr}\right)^2 + \frac{1}{r}\frac{d\rho}{dr} - \frac{\rho}{\varepsilon}\frac{\sqrt{\sigma_0^2 + B\rho^2}}{D} = 0 \tag{31}$$

With

$$B = \left(\frac{Dez}{kT}\right)^2 \tag{32}$$

No analytical solution is known for equation (31) so it has to be solved numerically for example by using a Runge Kutta method [16]. In this resolution, limit conditions for space charge density must be known. At r = 0, the assumption $\rho(0) = 0$ is made and at r = $R_d$, $\rho(R_d) = \rho_w$. Wall space charge density could be obtained by fitting in the model with no simplifying assumption with experimental values of the electrification like in the case of the weak space charge density assumption. To solve the equation (31), two boundary conditions need to be known, $\rho(R_d) = \rho_w$ and $\left(\frac{d\rho}{dr}\right)_{r=R_d}$. When a value of the wall space charge $\rho_w$ is chosen, $\left(\frac{d\rho}{dr}\right)_{r=R_d}$ is adjusted until this value leads to a $\rho(r)$ function which limit is zero for r=0. Then the $\rho_w$ value is adjusted until the calculated amount of charges convected by the flow, defined by equation (35) is equal to the experimental value.

By assuming that $\varepsilon_r = 2.7$, $\sigma_0 = 1.10^{-12}$ S/m and $\rho_w = 20$ C/m$^3$ which are typical values for polymer materials, one can calculate the space charge profile within a capillary of $R_d = 0.25$ mm (figure 3). The determination of $\rho_w$ value is essential to calculate the charge profiles thanks to equation (30) or equation (31) by the Runge Kutta method.

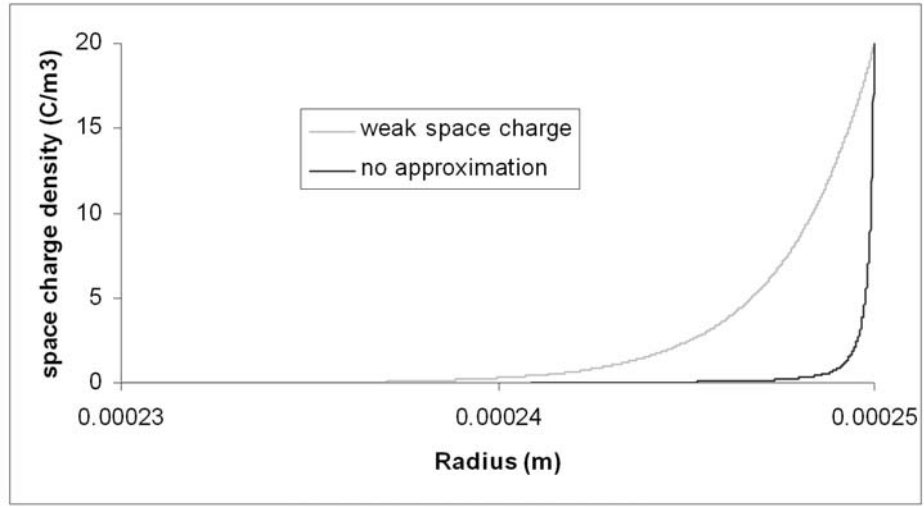

**Figure 3.** Space charge profile for weak space charge and without approximation

Figure 3 shows that the double layer extends over some hundred of micrometers. The double layer is localized very close to the capillary wall. Note also in figure 3 that the calculation approximations have a strong incidence on the space charge profile for polymers. The weak space charge approximation gives results very overestimated compared to the results obtained without approximation. So one first important result is that no approximation can be done while solving the equation (31) for polymer materials opposite from what is generally done for other types of liquids. One can notice that Taylor has used the weak space charge assumption for its polymers [5].

*3.3. Velocity profile inside the fluid flow*
As previously explained, after the determination of the charge density in the die, the second main component to evaluate the generated electric charges at the outlet of the die is the fluid velocity profile. Usually, for aqueous or low viscosity liquids, scientists use a Newtonien velocity profile, coming from the Newtonian law η=cst, in laminar condition [1, 19].

$$V(r) = \frac{\Delta P}{4L\eta}(R_d^2 - r^2) \quad (33)$$

where ΔP is the pressure gradient within the die, L is the length of the die and η is the polymer viscosity.

In the only study devoted to the particular case of flow electrification of polymer materials, Taylor used a non Newtonian velocity profile that looks like the velocity profile obtained from Ostwald law (see equation (1)) [5]. It is easy to demonstrate that Ostwald velocity profile can be written:

$$V(r) = R_d^{1/n+1} \left(\frac{\Delta P}{2KL}\right)^{1/n} \left(\frac{1}{1+1/n}\right) \left(1 - \left(\frac{r}{R_d}\right)^{1+1/n}\right) \quad (34)$$

where K and n are the fitted parameters of the Ostwald law. But we previously demonstrated in § 2.2 that for low shear rates this law does not lead to good agreements with the experimental results and so we proposed, for the studied polymers to use the Bird-Carreau law instead of the Ostwald law. Note that a great number of polymer behaviors can be well described by the Bird-Carreau law, and that our results can then be generalized to most polymers. In the last case of the Bird-Carreau law, the velocity profile can not be determined analytically but it can be easily obtained using a commercial finite elements simulation package like POLYFLOW. A comparison between a Newtonian and a non Newtonian velocity profile from Bird-Carreau law is shown in figure 4 for PDMS B at T = 60°C.

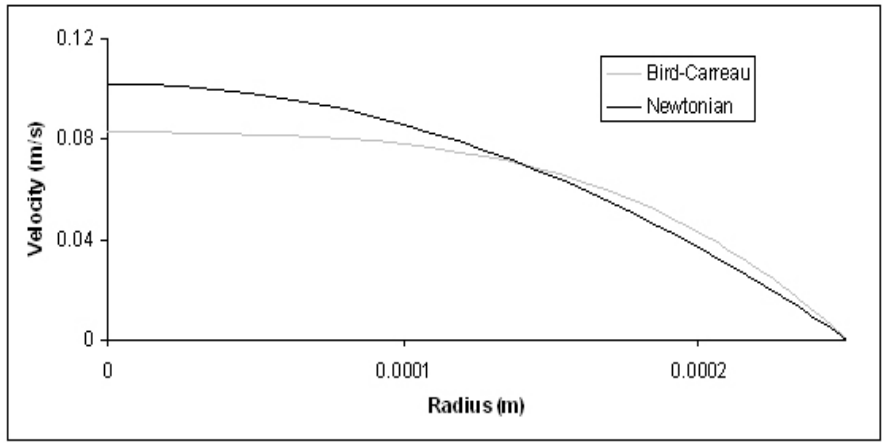

**Figure 4.** Comparison between two velocity profiles obtained respectively with a Newtonian law and a Bird-Carreau law at a fixed shear rate (800 s$^{-1}$)

An important consequence of the use of a Bird-Carreau law is the flattening of the velocity profile in comparison with a Newtonian velocity profile. This means that the fluid velocity is higher close to the wall where the electrical double layer is localized (figure 3) and that the convected charge will be more important in the case of non Newtonian velocity profile than for a Newtonian velocity profile.

Polymer is a highly viscous material with a low thermal conductivity. All of this results in a large amount of self heating that cannot be diffused cross flow and leads to very high and thin peaks of temperature. Moreover, polymer viscosity is quickly diminishing with temperature [20]. So the assumption made to obtain equation (26) must be reconsidered in this case. Calculations were performed with POLYFLOW by taking into account the viscous heating. Temperature increases obtained at different flow rate are reported in table 3.

**Table 3.** Anisothermal calculation results

| Shear rates (s$^{-1}$) | 80 | 800 | 8000 |
|---|---|---|---|
| ΔT (K) | 0.02 | 0.6 | 9 |

In the range of shear rates used in this work (80 to 8000 s$^{-1}$) the increases of the temperature due to viscous heating can be neglected. But, for higher shear rates, the viscous heating would be considered in the calculation of the velocity profile and also of the space charge density profile.

Another important aspect concerning the velocity profile is the occurrence of slip at the liquid/die wall interface. Slip velocity occurrence change the velocity profile and will induce a dramatic increase of the electrical charging. In the particular case of polymers, many teams have reported the presence of such a slip velocity for every flow rate [21-23]. We think that this parameter is one major key to understand the flow electrification of polymer materials.

Later in this article, we will discuss about the influence of such velocity profiles on flow electrification.

*3.3 Total space charge convected by the flow*

When the velocity profile and the charge profile are determined, the total space charge density can be calculated thanks to the following equation:

$$Q_\infty = \frac{2}{R_d^2 V_{mean}} \int_0^{R_d} \rho(r) \times V(r) r \, dr \qquad (35)$$

With $Q_\infty$ the total space charge for fully developed double layer, $V_{mean}$ the main velocity, $\rho(r)$ the space charge density at radius r and $V(r)$ the velocity at radius r.

For short dies or insulating materials, the double layer can be not fully developed. The time necessary to obtain a fully developed double layer is given by:

$$\tau = \frac{\varepsilon}{\sigma} \qquad (36)$$

In our case, this time varies approximately from 10s to 4ms. On the other hand, polymer residence time in the capillary is about 2.4s to 32ms. So the double layer will be rarely fully developed and length effect must be taken into account [5].

This is easily accomplished by using the following equation (37),

$$Q_L = Q_\infty \left[ 1 - \exp\left( - \frac{L}{V_{mean} \tau} \right) \right] \qquad (37)$$

Where $Q_L$ is the total space charge density for a given die length (L) and $\tau$ is the dielectric relaxation time. Such an effect will be always considered in the following parts of this work.

**4. Discussion**

This part will be devoted to apply the above theory to experimental cases. Firstly, we will discuss about the influence of the velocity profile on the total space charge density convected by the flow. Secondly, the wall space charge density parameter will be presented and we will discuss about its importance. Lastly, theoretical calculations of the space charge density convected by the flow considering a non zero velocity at the wall will be discussed.

*4.1. Influence of the velocity profile*

To observe the velocity profile influence on the electrical charging we consider three different profiles. The first one is the Newtonian velocity profile obtained thanks to the equation (33) and corresponding to standard liquids in laminar conditions. The second one is a non Newtonian velocity profile obtained from the Bird-Carreau law and evaluated by simulation with POLYFLOW. The last one is a flat velocity profile with a non zero velocity value at the wall (plug flow). To calculate the total space charge density convected by the flow, we use the same space charge density profile as in figure 3. Results are reported in figure 5.

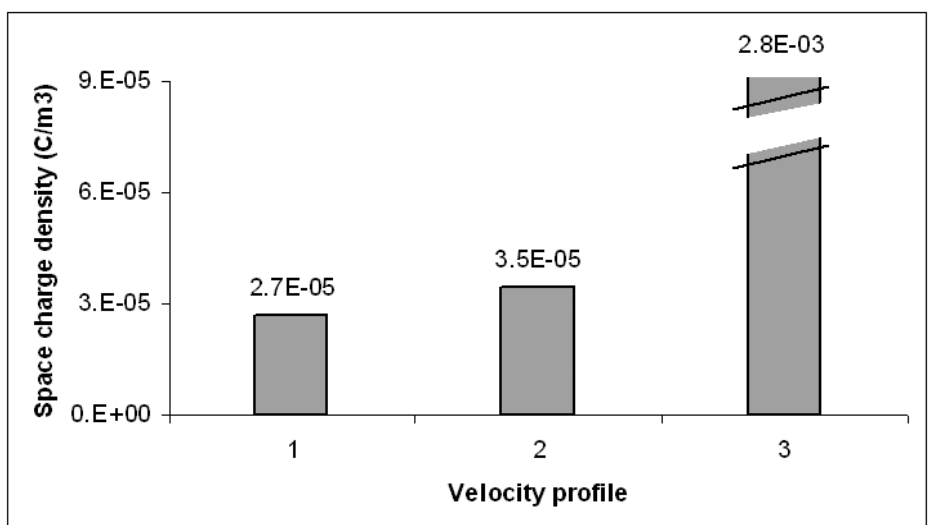

**Figure 5.** Space charges convected by the flow for different velocity profile for a given shear rate (80 s$^{-1}$): (1) Newtonian, (2) Bird-Carreau and (3) flat profiles.

In figure 5, one can observe important differences between the total space charges convected by the three velocity profiles even if the charge density profiles were the same. With the Bird-Carreau velocity profile, the total space charge density convected by the flow increases about 20% in comparison with the one obtained with the Newtonian velocity profile for the same space charge profile. This result underlines the importance of the choice of the velocity profile to modelise the flow electrification of polymers. Another important result from figure 5 is the huge increase of the total space charge density convected by the flow in the case of a flat velocity profile (multiplication by 100 between the Newtonian profile and the flat profile). These two increases can be easily explained by the rise of the velocity close to the die wall where the maximum charge density is. This maximum near the die wall and the quick decrease of charge density along the capillary radius also suggest that the occurrence of a slip velocity at the wall could strongly influence the flow electrification of polymers.

*4.2. Evaluation of the space charge density at the wall*
To calculate the space charge profiles of the figure 3, we used some data concerning the dielectric liquid like its electrical properties (permittivity and conductivity given in table 2) and some other experimental values. We also made the assumption that the wall space charge density, $\rho_w$, was equal to 20 C/m$^3$. This parameter, introduced by Touchard, represents the chemical interaction between the electrical charges present in the liquid and the die material [1]. This means that $\rho_w$ changes from one polymer to another or for one die material to another or also with the temperature. But for a given polymer, a given temperature and a fixed die this parameter does not change whatever the applied shear rate. However no theory can yet predict the value of this parameter and it must be determined experimentally as shown above.

Starting from the experimental value of the space charge convected by the flow as shown in Figure 6a (and measured at the outlet of the die) and from the knowledge of the velocity profile for a given flow rate, the determination of the wall space charge density is accomplished by adjusting it until the calculated convected space charge density (at the outlet of the die) corresponds to the measured ones (§3.2). This work has been done for example for PDMS A for a 30/1 die at T = 80°C. First, the Newtonian assumption has been made in the calculation of the velocity profile because the proposed range of shear rates is neighboring the Newtonian plateau (figure 1). Anisothermal effects were neglected. $\rho_w$ results are reported in figure 6b and the corresponding calculated curve of the space charge density is plot on figure 6a. With such a procedure the solution for $\rho_w$ is unique for a given shear rate.

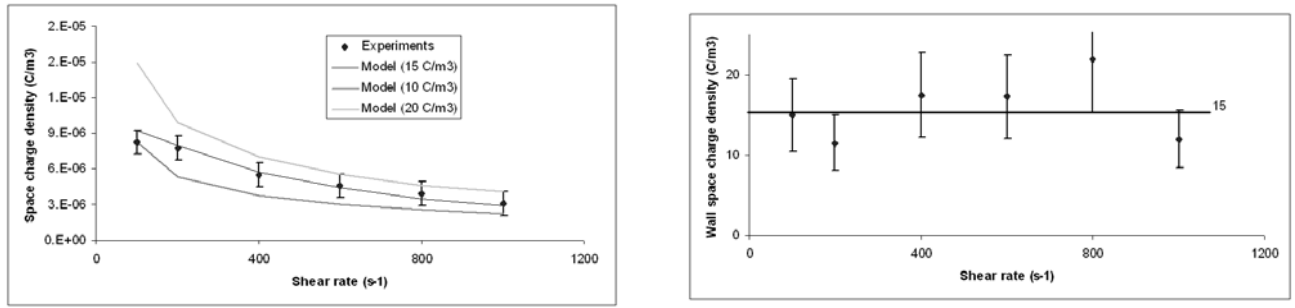

**Figure 6.** a) Measured total convected charge versus shear rate; the plain curves are the calculated values obtained for different wall space charge density values. b) Wall space charge density versus shear rate determined from experimental values of the total convected charge shown in a).

According to the figure 6, there is a mean value for the wall space charge density $\rho_w$ around 15 C/m$^3$. The fact that the wall space charge density is constant whatever the shear rate is consistent with the theory described above. Moreover, that also implies that a Newtonian velocity profile gives good electrification results in this low shear rate range for PDMS A. Another calculation performed with Bird-Carreau velocity profile gives very similar results for wall space charge density $\rho_w$. This likeness of the results obtained from Newtonian or Bird-Carreau laws is normal because for such a temperature (T = 80°C) and such shear rates lower than 1000 s$^{-1}$ the Bird-Carreau law is in good accordance with the Newtonian law.

To check if the theory also gives good agreements for higher shear rates and lower temperature, we have performed experiments with PDMS B for 40/0.5 die at T = 60°C. For such temperature (T = 60°C) the viscosity plateau is high around 350 Pa.s. Results are reported in figure 7.

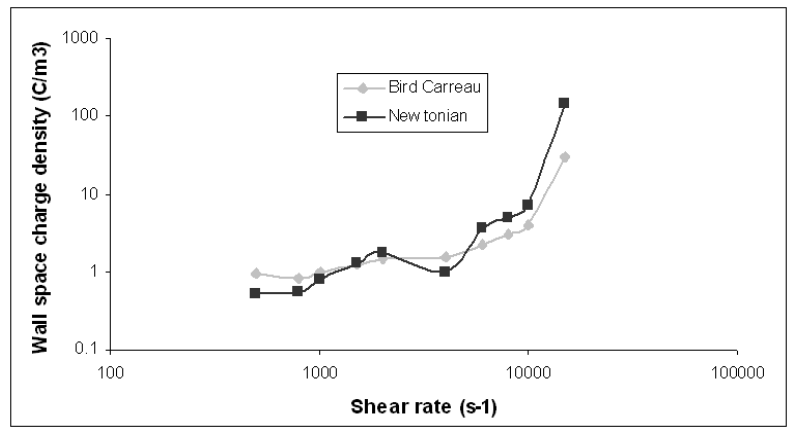

**Figure 7**. Wall space charge density versus shear rate.

Below 1000 s$^{-1}$, wall space charge density is quite constant and between 1000 s$^{-1}$ and 4000 s$^{-1}$ one can observe a modest increase of $\rho_w$. But above 4000 s$^{-1}$ a surprising feature of the experimental result is the increase of the wall space charge density with the shear rate. In § 2.2, we proved that polymer are non Newtonian fluids materials and their behavior (shear-thinning) could be modelize by the Bird-Carreau law. Furthermore, in § 3.3, we have put into evidence that the velocity profile change from Newtonian behavior to non Newtonian behavior. Lastly, we showed in § 4.1 that the velocity profile induces important changes on the convected total space charge. So, one way to eliminate this discrepancy can be to consider a better adapted velocity profile instead of the Newtonian one. Calculations, thanks to Bird-Carreau velocity profile, are also reported on the figure 7. With the Bird Carreau profile and for shear rates below 4000 s$^{-1}$ wall space charge density is quite constant. That is consistent with the fact that the Bird-Carreau law is the best adapted law to describe the behavior of this particular polymer at this temperature. For higher shear rates, $\rho_w$ increases with shear rate in the same manner than for a Newtonian velocity profile. So, taking into account the non Newtonian behavior of the polymer flow could be an answer to modelize electrification at low shear rate for viscous polymers. But some discrepancies at high shear rates remain misunderstood.

On one hand, it is well-known that the velocity profile of polymer could change at high shear rates, by the onset of a slip at the polymer/die interface [24]. On the other hand, as shown previously in § 4.1 with the results concerning the flat velocity profile, a way to obtain a high electrification of the extrudate is to increase the velocity of the fluid in the area where the charge density due to the double layer is high i.e close to the polymer/die interface. Taking into account a slip velocity at the die wall can then be a coherent explanation of this increase of the total charge density measured on the extruded polymer at high shear rates. Consequently, a new approach is needed to calculate the total space charge density convected by the flow by introducing a slip velocity at the wall.

*4.3. Calculation of the space charge density convected by the flow with slip velocity*

In this particular case, as a first assumption, the velocity can be divided in two terms as shown in the literature [9]:

$$V(r) = V(r)_{noslip} + V_s \qquad (38)$$

where $V(r)_{noslip}$ is the velocity without slip and $V_s$ is the slip velocity. The corresponding velocity profile with no-slip is calculated thanks to Newtonian assumption (equation (33)), and is adjusted so that the flow rate deduced from equation (38) corresponds to the experimental ones.

Then equation (35) becomes,

$$Q_\infty = \frac{2}{R_d^2 V_{mean}} \int_0^{R_d} \rho(r) V(r)_{noslip} r\, dr + \frac{2 V_s}{R_d^2 V_{mean}} \int_0^{R_d} \rho(r) r\, dr \qquad (39)$$

From equation (39) it is seen that the total space charge convected by the flow can be divided in two terms. The first one is the classical expression presented above. The second one is only due to the slip velocity. It is easy to understand that for the same wall space charge density, the total space charge convected by the flow will be higher. Respectively, if the total space charge convected by the flow is kept constant, the higher the slip velocity, the lower the wall space charge density. So, taking into account the slip velocity is an answer to eliminate the discrepancies observed in figure 7.

Inversely a slip velocity can easily be put in evidence and calculated by differentiating the total space charge convected by the flow with the total space charge convected by the velocity profile without slip to keep constant the wall space charge density. This is an important fact because this mean that we are able to calculate slip velocity of a polymer thanks electrification. This work has been performed and results for the PDMS B at 60°C are plotted on figure 8.

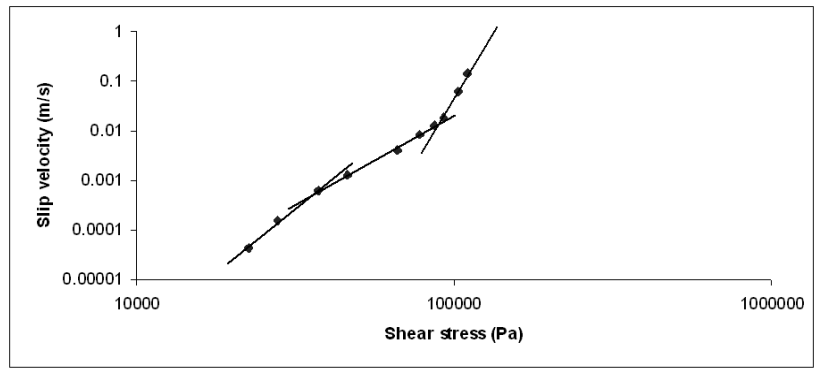

**Figure 8**. Slip velocity versus wall shear stress for PDMS B at T = 60°C

For this polymer, the slip velocity increases with the wall shear stress from 3 μm/s to 0.1 m/s. So calculation of the slip velocity thanks electrification measurements alow to obtain values in a large range of slip behavior from weak slip to strong slip. These orders of magnitude are consistent with other slip velocity values obtained with others experimental setups like Laser Doppler Velocimeter [25]. This last experimental set up allows performing direct measurements of slip velocity near the die wall, at some tens of microns from the die wall. Work is still in progress on the use of the electrification measurement to evaluate slip velocity at the polymer/die wall interface, it will be detailed in a future paper. The slip results obtained with this method will be accurately compared with other published results on wall slip.

## 5. Conclusion

In this work we studied the flow electrification of polymers through capillaries. The electrical charging of the extrudate can be directly measured at the outlet of the die during the extrusion process. To explain and quantify this electrification phenomenon, we adapted the double layer theory to the polymer case.

We first demonstrated that the solving of the transport equations necessary to evaluate the charge profile has to be performed without approximation contrary to the case of usual aqueous liquids or oils.

To obtain the total charge convected by the flow we also need to know the fluid velocity profile. As previously shown by Taylor, we confirmed that the non Newtonian nature of the polymer flow has a strong influence on the electrification evaluation. Indeed the flow velocity profile has to be deduced from viscosity laws better adapted to the polymer case than the simplest Ostwald ones.

Finally we showed that taking a non-zero velocity at the wall into account is a consistent way to explain some unexpected results such as an apparent non-constant wall charge density. The slip velocity induces dramatical increases of flow electrification when it occurs. This new approach to modelise the polymer flow electrification allowed us to establish a strong link between the generated electrical charges and the occurrence of slip velocity.

Reciprocally, this electrification model can be used for slip velocity calculation in the case of polymer flow. One can note that this slip phenomenon can be weak thus difficult to quantify and it also usually needs a complex measurement set-up. Here, we proposed a simple device to evaluate slip velocity with a good resolution

## 6. Acknowledgements

The authors are greatly indebted to Pr G Touchard of the University of Poitiers for fruitful discussions on the double layer theory.